\documentclass[%
 reprint,
 amsmath,amssymb,
 aps,
]{revtex4-2}
\usepackage{graphicx}
\usepackage{dcolumn}
\usepackage{bm}
\usepackage{hyperref}
\usepackage{float}

\begin{document}

\preprint{APS/123-QED}

\title{Singly heavy Omega Baryon ($\Omega_c^0$ \& $\Omega_b^-$) Spectroscopy in the Relativistic Framework of Independent Quark Model}

\author{Rameshri V. Patel}
\email{rameshri.patel1712@gmail.com}
 \affiliation{P D Patel Institute of Applied Sciences, Charusat University, Anand 388421, Gujarat, India.}
\author{Manan Shah} 
\email{mnshah09@gmail.com}
\affiliation{P D Patel Institute of Applied Sciences, Charusat University, Anand 388421, Gujarat, India.}
\author{Smruti Patel}
\affiliation{Government Arts, Commerce, and Science College, Limbayat, Surat, Gujarat.}
\author{Bhoomika Pandya}
\affiliation{Department of Physics  Education, Teachers College, Kyungpook National University, Daegu 41566, Republic of Korea},\affiliation{Department of Physics, Education Research Center for Quantum Nature of Particles and Matter, Daegu 41566, Republic of Korea}

\date{\today}

\begin{abstract}
The Independent Quark Model, formulated for a three-body system within a relativistic framework, is applied to singly heavy baryons \( \Omega_c^0 \) and \( \Omega_b^- \) to investigate their spectroscopic properties. A Martin-like potential with an equal mixture of scalar and vector components is employed, with potential parameters fitted using ground-state experimental data. The resulting mass spectra include both radial and orbital excitations. The spin-parity for the recently observed states \( \Omega_c^0(3000) \), \( \Omega_c^0(3050) \), \( \Omega_c^0(3067) \), \( \Omega_c^0(3120) \), and \( \Omega_c^0(3185) \) as well as possible spin assignments for the four newly observed excited states of \( \Omega_b^- \) (\(\Omega_b^-(6316)\), \(\Omega_b^-(6330)\), \(\Omega_b^-(6340)\), and \(\Omega_b^-(6350)\)) are proposed. The magnetic moments of the ground and first excited states are also calculated, along with radiative decay widths and transition magnetic moments. The non-leptonic weak decays of \( \Omega_c^0 \) are analyzed, with decay widths and branching ratios computed and compared with experimental data to validate the predictive power of the model. The branching ratios for the non-leptonic decays of $\Omega_b^-$ are also predicted for future observations.    
\end{abstract}

\keywords{Baryon Spectroscopy, Phenomenology, Independent Quark Model, Mass Spectra, Power-law potential}
                              
\maketitle


\section{Introduction}
Heavy hadron spectroscopy has recently become more important and interesting due to the discovery of many new states. The classification of baryons with a single heavy quark ($Qqq'$)—where $Q$ is a heavy quark like bottom ($b$) or charm ($c$) and $q$, $q'$ are lighter quarks like up ($u$), down ($d$), or strange ($s$)—is effectively done using the heavy quark-diquark concept. Heavy-Quark Effective theory ($HQET$) helps simplify the study of these baryons by isolating the heavy-quark dynamics with corrections that are proportional to the inverse of the heavy-quark mass ($1/m_Q$). Observing new baryons and measuring their properties provide insights into the role of diquarks within baryons, and this information can also refine models of more complex states like tetraquarks and pentaquarks. Besides the challenges of exotic particles, many states are related to the excited states of known hadrons or mixtures of nearby states. For a long time, only two ground states had been observed for the $\Omega_c$ baryons: $\Omega_c^{0}$ ($J^P = 1/2^+$) and $\Omega_c(2770)^{0}$ ($J^P = 3/2^+$). The latter was detected by BABAR through electromagnetic decay $\Omega_c(2770) \to \Omega_c \gamma$ \cite{BaBar:2006pve}. The $\Omega_c^{0}$ baryon has become especially interesting after new resonances were found in the decay $\Omega_c^0 \to \Xi_c^+ K^-$\cite{LHCb:2017uwr}. These new states are $\Omega^{0}_c(3000)$, $\Omega^{0}_c(3050)$, $\Omega^{0}_c(3067)$, $\Omega^{0}_c(3090)$, and $\Omega^{0}_c(3120)$. The LHCb collaboration also suggested possible spins for these states and published their findings in 2021\cite{LHCb:2021ptx}. However, the very recent observation of $\Omega^{0}_c(3185)$ and $\Omega^{0}_c(3327)$ made it necessary to study and apply the models to these baryons to understand the internal dynamics of the same\cite{LHCb:2023sxp}. 
The simplest explanation for the recent observation of four narrow peaks in the $\Xi_b^0K^-$ mass spectrum at LHCb are excited states of $\Omega_b^-$      \cite{LHCb:2020tqd}. These are likely $L=1$ orbital excitations of the ground state or possibly $n=2$ radial excitations \cite{LHCb:2020tqd}.

Over the past $27$ years, singly heavy baryons, such as $\Omega_c^0$ and $\Omega_b^-$, have been the subject of extensive theoretical investigations \cite{Chiladze:1997ev, Garcilazo:2007eh, Ebert:2007nw, Valcarce:2008dr, Roberts:2007ni, Wang:2007sqa, Liu:2007fg, Wang:2009ozr, Wang:2009cr, Ebert:2011kk, Vijande:2012mk, Padmanath:2013bla, Yoshida:2015tia, Chen:2015kpa, Aliev:2015qea, Lu:2014ina, Shah:2016nxi, Karliner:2017kfm, Chen:2017sci, Azizi:2015ksa, Kim:2017khv, Wang:2017kfr, Wang:2017hej, Cheng:2017ove, Zhao:2017fov, Wang:2017vnc, Padmanath:2017lng, Wang:2017zjw, Agaev:2017jyt, Agaev:2017lip, Agaev:2017ywp, Chen:2017gnu, An:2017lwg, Mao:2017wbz, Yao:2018jmc, Yang:2017rpg, Chen:2017xat, Nieves:2017jjx, Huang:2018wgr, Chen:2018vuc, Montana:2017kjw, Debastiani:2017ewu, Santopinto:2018ljf, Cui:2019dzj, Yang:2019cvw, Xu:2019kkt, Ramos:2020bgs, Hu:2020nkg, Yang:2021lce, Wang:2021cku, RamirezMorales:2022lsy, Kakadiya:2022zvy, Jakhad:2023mni, Oudichhya:2023awb, Oudichhya:2024eax, Jakhad:2024fin, Yu:2022ymb}. Various frameworks, including relativistic quark models, the QCD sum rule method, heavy-quark effective theory, diverse quark models, and lattice quantum chromodynamics (QCD) simulations, have been employed to study their fundamental properties. These efforts have primarily focused on predicting their masses, strong couplings, transition amplitudes, magnetic moments, radiative decay rates, and a variety of other decay characteristics \cite{Chiladze:1997ev, Garcilazo:2007eh, Ebert:2007nw, Valcarce:2008dr, Roberts:2007ni, Wang:2007sqa, Liu:2007fg, Wang:2009ozr, Wang:2009cr, Ebert:2011kk, Vijande:2012mk, Padmanath:2013bla, Yoshida:2015tia, Chen:2015kpa, Aliev:2015qea, Lu:2014ina, Shah:2016nxi, Karliner:2017kfm, Chen:2017sci, Azizi:2015ksa, Kim:2017khv, Wang:2017kfr, Wang:2017hej, Cheng:2017ove, Zhao:2017fov, Wang:2017vnc, Padmanath:2017lng, Wang:2017zjw, Agaev:2017jyt, Agaev:2017lip, Agaev:2017ywp, Chen:2017gnu, An:2017lwg, Mao:2017wbz, Yao:2018jmc, Yang:2017rpg, Chen:2017xat, Nieves:2017jjx, Huang:2018wgr, Chen:2018vuc, Montana:2017kjw, Debastiani:2017ewu, Santopinto:2018ljf, Cui:2019dzj, Yang:2019cvw, Xu:2019kkt, Ramos:2020bgs, Hu:2020nkg, Yang:2021lce, Wang:2021cku, RamirezMorales:2022lsy, Kakadiya:2022zvy, Jakhad:2023mni, Oudichhya:2023awb, Oudichhya:2024eax, Jakhad:2024fin, Yu:2022ymb}. Despite these extensive studies, no approach has comprehensively and accurately predicted the spin-parity assignments of all recently observed excited states of these baryons. Recognizing this significant gap in understanding their properties, we investigate their mass spectra, magnetic moments, and decay characteristics, aiming to provide a deeper insight into these systems and assess the predictive success of our theoretical model.

 The Independent Quark Model (IQM) was initially developed by A. Kobushkin \cite{Kobushkin:1976fq} and P. Ferreira \cite{LealFerreira:1977gz} to describe the linear confinement of quarks. Their framework proposed that individual quarks within a baryon obey a Dirac-type equation, with an average potential centered around the hadron's center of mass. In later developments it was  demonstrated that expressing the average potential as an equal mix of scalar and vector components simplifies calculations. This transformation turns the single-quark Dirac equation into an effective Schrödinger equation \cite{Barik:1982nr, Barik:1985rm, Barik:1986gw}.
The study of quark confinement within a baryon can be carried out using the Martin-like potential, which includes an equal combination of scalar and vector components. This potential has been applied in the relativistic context of the Independent Quark Model (IQM) for various mesons \cite{Shah:2014caa, Vinodkumar:2014afm, Shah:2014aly, Vinodkumar:2015blb, Shah:2014yma, Shah:2016mgq,Pandya:2021ddc}. Due to its favorable results and effectiveness in predicting and validating experimental observations for mesons, we recently adapted and refined this model for the use of different types of baryons \cite{Patel:2023wbs, Shah:2023mzg}.

This paper presents a comprehensive discussion of the general methodology in Sec. \ref{sec:2}, designed to be applicable to any baryonic system. The outlined approach involves solving the Dirac equations for individual quarks within a baryon, enabling the determination of spin-averaged masses and contributions arising from spin-spin, spin-orbit, and tensor interactions. Furthermore, we calculated key static properties, including the magnetic moment along with radiative decay width of a baryon, comparison of magnetic moment with the experimental results offers a critical test of our model's validity, which is provided in the Sec. \ref{sec:3}. The calculation for the non-leptonic weak decays of $\Omega_c^0$ and non-leptonic decays of $\Omega_b^-$. The detailed conclusion is presented in Sec. \ref{sec:4}

\section{Methodology}\label{sec:2}
The independent quark model was developed in detail for the system of two quarks, i.e., mesons \cite{Shah:2014caa, Vinodkumar:2014afm, Shah:2014aly, Vinodkumar:2015blb, Shah:2014yma, Shah:2016mgq,Pandya:2021ddc}.
We have developed it further to include three quark systems \cite{Patel:2023wbs, Shah:2023mzg, Patel:2024wfo,Patel:2023djz}.
In this model, the properties of individual quarks are governed by a Dirac equation formulated in the hadron's rest frame. The potential in this equation showcases a Lorentz structure featuring an even blend of scalar and vector components.
The initial step in this formalism is to calculate the spin average masses of the multiple quark systems; for the baryons, which can be calculated as
\begin{eqnarray}\label{sa}
    M_{SA}^{q_1q_2q_3} = E_{q_1}^D+E_{q_2}^D+E_{q_3}^D-E_{CM}.
\end{eqnarray}
Here, $E_{CM}$ incorporates the center of mass corrections removing all the irrelevant effects from the dynamics of the center of mass. $E_{q_i}^D$ ($i=1,2,3$) represents the Dirac energies of the individual quark in the baryon system. These energies calculated by solving the Dirac equation for a quasi-independent quark in the center of mass frame that has the form of,
\begin{equation}\label{dirac}
    \left[E^D_q - \boldsymbol{\hat{\alpha}} .\boldsymbol{\hat{p}} - \hat{\beta} m_q - V(r)\right]\psi_q(\vec{r}) = 0,
\end{equation}
where $E^D_q$ represents the Dirac energy of a quark, $m_q$ is the current quark mass and $\psi_q(\vec{r})$ is the four-component quark wave-function which is a spinor.
Quarks within a hadron are hypothesized to move independently under the influence of a flavor-independent central potential. The central potential is described by a Martin-like functional form given as  
\begin{equation}\label{potential}
    V(r) = \frac{(1+\gamma_0)}{2}(\Lambda r^{0.1} + V_0),
\end{equation}  
where $\Lambda$ represents the strength of the potential, and $V_0$ denotes its depth. As discussed in \cite{Greiner:2000cwh}, the solution of Eqn.(\ref{dirac}) can be expressed as 
\begin{equation}
    \psi_q(\vec{r}) = \begin{pmatrix}
    i g(r) \Omega_{jlm} \left(\frac{\boldsymbol{r}}{r}\right)\\
    -f(r) \Omega_{jl'm} \left(\frac{\boldsymbol{r}}{r}\right)
    \end{pmatrix},
\end{equation}
for the potential of Eqn.(\ref{potential}). Here, the spinor spherical harmonics $\Omega_{jlm}$ defined as given in Ref. \cite{Greiner:2000cwh},
\begin{equation}
    \Omega_{jlm} = \sum_{m',m_s} \big(l\frac{1}{2}j \rvert m'm_s m\big) Y_{lm'} \chi_{\frac{1}{2}m_s},
\end{equation}
with parity $\hat{P}_0 \Omega_{jlm} = (-1)^l \Omega_{jlm}$, $\chi_{\frac{1}{2}m_s}$ being eigenfunctions of $\boldsymbol{\hat{S}}^2$ $\&$ $\hat{S}_3$ and $Y_{lm'}$ being the spherical harmonics. 
In order to estimate the Dirac energies, we need to mathematically rearrange the equations for $f(r)$ \& $g(r)$ to make them equivalent to ODE obeyed by the reduced radial part of the Schrödinger wave function \cite{Barik:1982nr}    
\begin{multline}\label{r}
        \frac{d^2R^{Sch}(r)}{dr^2} + \Bigg[m_q(E^{Sch}_q-V(r))\\-\frac{l(l+1)}{r^2}\Bigg]R^{Sch}(r) = 0.
\end{multline}
The radial parts of Dirac spinors follow
second-order ordinary differential equations.
    \begin{multline}\label{g}
        \frac{d^2g(r)}{dr^2} + \Bigg[(E^D_q+m_q)[E^D_q-m_q-V(r)]\\-\frac{k(k+1)}{r^2}\Bigg]g(r) = 0,
    \end{multline}
    \begin{multline}\label{f}
        \frac{d^2f(r)}{dr^2} + \bigg[(E^D_q+m_q)[E^D_q-m_q-V(r)]\\-\frac{k(k-1)}{r^2} \bigg]f(r) = 0,
    \end{multline}
where $k$ is the eigenvalue of the operator $\hat{k} = (1+\boldsymbol{\hat{L}}\cdot\boldsymbol{\hat{\sigma}})$ having the value, 
\begin{equation}
    k = \begin{cases} 
            -(l+1) = -\left(j+\frac{1}{2}\right) \hspace{0.7cm} for \hspace{0.1cm} j = l+\frac{1}{2}\\ \hspace{1.15cm}l = +\left(j+\frac{1}{2}\right) \hspace{0.7cm} for \hspace{0.1cm} j = l-\frac{1}{2}  
        \end{cases}
\end{equation}
For the potential of Martin-like for Eqn.(\ref{potential}), these ODEs can be made equivalent to ODE obeyed by the reduced radial part of the Schrödinger wave function by defining a dimensionless variable $\rho = \frac{r}{r_0}$ that will reduce these Eqns. (\ref{r}), (\ref{g}) \&(\ref{f}) to the equivalent form,
\begin{equation}\label{grho}
\frac{d^2g(\rho)}{d\rho^2} + \left[\epsilon^{D} - \rho^{0.1} - \frac{k(k+1)}{\rho^2}\right]g(\rho) = 0,
\end{equation}
\begin{equation}
\frac{d^2f(\rho)}{d\rho^2} + \left[\epsilon^{D} - \rho^{0.1} - \frac{k(k-1)}{\rho^2}\right]f(\rho) = 0,
\end{equation}
\begin{equation}
\frac{d^2R^{Sch}(\rho)}{d\rho^2} + \left[\epsilon^{Sch} - \rho^{0.1} - \frac{l(l+1)}{\rho^2}\right]R^{Sch}(\rho) =0.
\end{equation}
Here, 
\begin{equation}
\epsilon^D = (E^D_q - m_q - V_0)(m_q + E^D_q)^\frac{0.1}{2.1}\left(\frac{2}{\Lambda}\right)^\frac{2}{2.1}
\end{equation} 
and
\begin{equation}
\epsilon^{Sch} = m_q\left(E^{Sch}_q - V_0\right)\left(m_q\right)^\frac{-2}{2.1}\left(\frac{1}{\Lambda}\right)^{\frac{2}{2.1}}.
\end{equation}
In the Schrodinger case, $r_0 = (m_q \Lambda)^{\frac{-1}{2.1}}$ and $r_0 = \left[(m_q + E^D_q)\frac{\Lambda}{2}\right]^{\frac{-1}{2.1}}$ in the Dirac case \cite{Barik:1982nr}. The Schrödinger equation can be solved numerically using the code given in Ref.\cite{W Lucha : 1998} and the Dirac energies for the individual quarks can be found by equating $\epsilon^D$ to $\epsilon^{Sch}$.

The parameters of the potential are determined by matching the theoretical spin-averaged mass, as given by Eq. (\ref{sa}), to the experimental spin-averaged mass of the $S$ wave. The experimental spin-averaged mass is calculated using the expression  
\begin{equation}
M_{SA} = \frac{\sum_J (2J+1)M_{nJ}}{\sum_J (2J+1)},  
\end{equation}  
 for the $S$ waves of the baryon which simplifies to $(M_{1/2} + 2M_{3/2})/3$. Once the parameters are fitted, the spin-averaged masses for the excited $S$ waves can also be evaluated.

To account for spin degeneracy, the spin-spin interaction is incorporated into $ M_{SA} $ by considering the total spin of the quark system, defined as $ \vec{J}_{3q} = \vec{J}_1 + \vec{J}_2 + \vec{J}_3 $. The spin-spin interaction is described by the expression  
\begin{equation}\label{vjj}
\big<V^{jj}_{q_1q_2q_3}(r)\big> = \sum_{i=1, i<k}^{i,k=3} \frac{\sigma \big<j_i.j_kJM \rvert \widehat{j_i}.\widehat{j_k}\rvert j_i.j_kJM\big>}{(E^D_{q_i} + m_{q_i})(E^D_{q_k} + m_{q_k})},  
\end{equation}  
where the interactions are expressed as the sum of the contributions from individual quark pairs. Here, $ \sigma $ represents the $ j\text{-}j $ coupling constant, which is determined by fitting to experimental data. The fitted values for the potential parameters, center-of-mass corrections, and the $ j\text{-}j $ coupling constant for the $ \Omega_c^{0} $ and $ \Omega_b^{-} $ baryons are summarized in Tables \ref{tab:1} and \ref{tab:2}.

\begin{table}[!tbh]
\begingroup
\caption{Fitted parameters for the $\Omega_c^{0}$} \label{tab:1}
\setlength{\tabcolsep}{2pt}
\renewcommand{\arraystretch}{1.5}
\begin{tabular}{c c}
\hline
\hline
Parameter & Value (With $5\%$ variation)\\
\hline
Depth of the potential ($V_0$) & $-1.011\pm 0.050$ $GeV$  \\
Potential strength ($\Lambda$)  & $1.250 \pm 0.062$ $GeV^{1.1}$ \\
Center of mass correction  ($E_{CM}$)  & $0.236 \pm 0.012$ $GeV$ \\
$j-j$ coupling constant ($\sigma$) & $0.075 \pm 0.004$ $GeV^{3}$\\
\hline
\hline
\end{tabular}
\endgroup
\end{table}

\begin{table}[!tbh]
\begingroup
\caption{Fitted parameters for the $\Omega_b^{-}$} \label{tab:2}
\setlength{\tabcolsep}{2pt}
\renewcommand{\arraystretch}{1.5}
\begin{tabular}{c c}
\hline
\hline
Parameter & Value (With $5\%$ variation)\\
\hline
Depth of the potential ($V_0$) & $-1.110 \pm 0.055$ $GeV$  \\
Potential strength ($\Lambda$)  & $1.450 \pm 0.072$ $GeV^{1.1}$ \\
Center of mass correction  ($E_{CM}$)  & $0.102 \pm 0.005$ $GeV$ \\
$j-j$ coupling constant ($\sigma$) & $0.069 \pm 0.003
$ $GeV^{3}$\\
\hline
\hline
\end{tabular}
\endgroup
\end{table}

To derive the masses of the $P$, $D$, and  $F$ states from the spin-average mass, we incorporate three interactions: spin-spin, spin-orbit, and tensor interactions. The spin-spin interaction term is defined in Eqn.(\ref{vjj}). However, based on a phenomenological current confinement model for gluons, researchers have derived a closed analytical expression for the confined gluon propagator (CGP) in coordinate space, specifically for small frequencies, using a translationally invariant ansatz \cite{Vinodkumar:1992wu}. Utilizing the CGP, the complete one-gluon exchange potential (COGEP) between quarks has been formulated through the Fermi-Breit formalism. This formulation provides a valuable framework for exploring hadron spectroscopy and hadron-hadron interactions. The spin-orbit and tensor interaction terms emerge as integral components of the COGEP \cite{Vinodkumar:1992wu}, which are also considered to be the sum of interactions between the pairs of quarks,
    \begin{multline} 
    V^{LS}_{q_1q_2q_3}(r) =  \frac{\alpha_s}{4} \sum_{i=1,i<k}^{i,k=3} \frac{N_{q_i}^2.  N_{q_k}^2}{(E^D_{q_i}+m_{q_i})(E^D_{q_k}+m_{q_k})} \frac{\lambda_i.\lambda_j}{2r}\\
    \otimes [[r \times (\widehat{p}_{q_i}-\widehat{p}_{q_k}).(\widehat{\sigma}_{q_i}+\widehat{\sigma}_{q_k})].[(D'_0(r)+2D'_1(r))]\\
    +[r \times (\widehat{p}_{q_i}+\widehat{p}_{q_k}).(\widehat{\sigma}_{q_i}-\widehat{\sigma}_{q_k})].[(D'_0(r)-D'_1(r))]],
    \end{multline}
    \begin{multline}
    V^{T}_{q_1q_2q_3}(r) =  -\frac{\alpha_s}{4} \sum_{i=1,i<k}^{i,k=3} \frac{N_{q_i}^2  N_{q_k}^2}{(E^D_{q_i}+m_{q_i})(E^D_{q_k}+m_{q_k})}\\ \otimes \lambda_i.\lambda_j \left(\left(\frac{D''_1(r)}{3}-\frac{D'_1(r)}{3r}\right)S_{{q_i}.{q_k}}\right).
    \end{multline}

    Where $\lambda_i .  \lambda_j$ represents the color factor of the baryon and  $S_{{q_i}.{q_k}} = [3(\sigma_{q_i}{\hat{r}})(\sigma_{q_k}{\hat{r}})-\sigma_{q_i}\sigma_{q_k}]$, the running coupling constant can be calculated as
\begin{eqnarray} 
\alpha_s = \frac{\alpha_s(\mu_0)}{1+\frac{33-2n_f}{12\pi} \alpha_s(\mu_0) ln\left(\frac{E_{q1}^D+E_{q2}^D+E_{q3}^D}{\mu_0}\right)}.
\end{eqnarray} 
Where $\alpha_{s}(\mu_{0}=1GeV)=0.6$
is considered in the present study.
 We keep the parametric form of the confined gluon propagators $(D_0 \ \&  \ D_1)$ as it is mentioned in Ref
 .\cite{Vinodkumar:1992wu}
 \begin{equation} 
    D_0(r) = \left(\frac{\alpha_1}{r}+\alpha_2\right)exp\left(\frac{-r^2c_0^2}{2}\right)
    \end{equation}
    \begin{equation}
    D_1(r) = \frac{\gamma}{r}exp\left(\frac{-r^2c_1^2}{2}\right)
    \end{equation}
with $\alpha_1 = 0.036$, $\alpha_2 = 0.056$, $c_0 = 0.1017$ $GeV$, $c_1 = 0.1522$ $GeV$ and $\gamma = 0.0139$. 
After deriving the wavefunction, the normalization constant $ N_{q_i} $ for the individual quark wavefunction can be determined. This enables the evaluation of $ \langle \psi | V^{LS} | \psi \rangle $ and $ \langle \psi | V^T | \psi \rangle $ for all permutations of $ q_1 $, $ q_2 $, and $ q_3 $. Summing these contributions across all permutations provides the total contribution for a given state. Including the spin-spin interaction terms in this total allows for the calculation of the masses of the corresponding $ P $, $ D $, and $ F $ states. We have mentioned our predictions, with corresponding experimental observation, and the other theoretical predictions for the $S$, $P$, and $D$ state masses of $\Omega_c^0$ in Tables \ref{tab:3}, \ref{tab:4}, and \ref{tab:5} respectively and the $S$, $P$, \& $D$ state masses of $\Omega_b^-$ in Tables \ref{tab:6}, \ref{tab:7} and \ref{tab:8} respectively.
\begin{table*}[ht]
{
\begingroup
\caption{$S$ State masses of $\Omega_c^{0}$(in $GeV$)} \label{tab:3}
\setlength{\tabcolsep}{5pt}
\renewcommand{\arraystretch}{1.5}
\begin{tabular}{c c c c c c c c c c }
\hline
\hline
$nL$ & $J^{P}$ & State & $\big<V^{jj}_{q_1q_2q_3}\big>$ & Our & Exp.\cite{ParticleDataGroup:2024cfk} & \cite{Shah:2016nxi} & \cite{Jakhad:2023mni} &   \cite{Yu:2022ymb}  & \cite{Ebert:2011kk}\\
\hline
$1S$ & $\frac{1}{2}^+$ & $1^2S_{\frac{1}{2}}$ & $-0.044$ & $2.698\pm0.048$ & $2.695\pm0.001$ &$2.695$&$2.695$ & $2.699$ & $2.698$ \\
$1S$ & $\frac{3}{2}^+$ & $1^4S_{\frac{3}{2}}$ & $0.026$ & $2.769\pm0.048$ & $2.765\pm0.002$  &$2.767$ & $2.766$ & $2.762$ &$2.768$ \\

$2S$ & $\frac{1}{2}^+$ & $2^2S_{\frac{1}{2}}$ & $-0.034$ & $3.047\pm0.064$ & $-$  &$3.100$ & $3.171$ & $3.150$ & $3.088$ \\
$2S$ & $\frac{3}{2}^+$ & $2^4S_{\frac{3}{2}}$ & $0.020$ & $3.102\pm0.064$ & $-$ &$3.126$&$3.180$ & $3.197$ & $3.123$ \\

$3S$ & $\frac{1}{2}^+$ & $3^2S_{\frac{1}{2}}$ & $-0.030$ & $3.248\pm0.073$ & $-$ &$3.436$ &$3.522$ & $3.308$ & $3.489$\\
$3S$ & $\frac{3}{2}^+$ & $3^4S_{\frac{3}{2}}$ & $0.018$ & $3.296\pm0.072$ & $-$ &$3.450$ &$3.524$ & $3.346$ & $3.510$ \\

$4S$ & $\frac{1}{2}^+$ & $4^2S_{\frac{1}{2}}$ & $-0.027$ & $3.392\pm0.079$ & $-$  &$3.737$&$-$ & $3.526$ & $3.814$ \\
$4S$ & $\frac{3}{2}^+$ & $4^4S_{\frac{3}{2}}$ & $0.016$ & $3.436\pm0.079$ & $-$  &$3.745$&$-$ & $3.557$ & $3.830$ \\

$5S$ & $\frac{1}{2}^+$ & $5^2S_{\frac{1}{2}}$ & $-0.025$ & $3.504\pm0.084$ & $-$ &$-$ & $-$ &$-$& $4.102$ \\
$5S$ & $\frac{3}{2}^+$ & $5^4S_{\frac{3}{2}}$ & $0.015$ & $3.545\pm0.084$ & $-$  & $-$& $-$ &$-$& $4.114$ \\
\hline
\end{tabular}
\endgroup}
\end{table*}
\begin{table*}[ht]
{
\begingroup
\caption{$P$ State masses $\Omega^0_c$(in $GeV$)} \label{tab:4}
\setlength{\tabcolsep}{5pt}
\renewcommand{\arraystretch}{1.5}
\begin{tabular}{ c c c c c c c c c c c}
\hline
\hline
$n^{2S+1}L_J$ & $\big<V^{jj}_{q_1q_2q_3}\big>$& $\big<V_{q_1q_2q_3}^{L.S}\big>$ & $\big<V_{q_1q_2q_3}^T\big>$& Our & Exp.\cite{ParticleDataGroup:2024cfk} & \cite{Yu:2022ymb}& \cite{Ebert:2011kk}& \cite{Shah:2016nxi} & \cite{Jakhad:2023mni}\\
\hline
$1^2P_{\frac{1}{2}}$ & $-0.039$ & $-0.003$ & $-0.001$ & $2.944 \pm 0.059$ & $-$ & $3.057$ & $3.055$ & $3.011$ & $3.003$\\
$1^2P_{\frac{3}{2}}$ & $0.025$ & $-0.0004$ & $0.0001$ & $3.012 \pm 0.059$ & $-$ & $3.062$& $3.054$ & $2.976$ &$3.063$\\
$1^4P_{\frac{1}{2}}$ & $-0.031$ & $-0.004$ & $-0.003$ & $2.949 \pm 0.058$ & $3.000\pm0.0002$ &$3.045$ & $2.966$ & $3.028$ &$3.011$\\
$1^4P_{\frac{3}{2}}$ & $-0.043$ & $-0.001$ & $0.001$ & $2.944 \pm 0.059$ & $-$ & $3.039$ & $3.029$ & $2.993$ &$3.107$\\
$1^4P_{\frac{5}{2}}$ & $0.057$ & $0.001$ & $-0.0002$ &$3.045 \pm 0.059$ & $3.050\pm0.0001$ & $3.067$ & $3.051$ & $2.947$ &$3.145$\\
\hline
$2^2P_{\frac{1}{2}}$ & $-0.032  $ & $-0.002$ & $-0.001$ & $3.177 \pm 0.070$ & $-$ &$3.426$& $3.435$ & $3.345$&$3.414$ \\
$2^2P_{\frac{3}{2}}$ & $0.020$ & $-0.0003$ & $0.00004$ & $3.233 \pm 0.069$ & $-$ & $3.431$ & $3.433$ &$3.315$& $3.437$\\
$2^4P_{\frac{1}{2}}$ & $-0.026$ & $-0.002$ & $-0.002$ & $3.182 \pm 0.070$ & $3.185\pm0.001$ &$3.416$& $3.384$ &$3.359$& $3.421$\\
$2^4P_{\frac{3}{2}}$ & $-0.034$ & $-0.001$ & $0.001$ & $3.179 \pm 0.070$ & $-$ & $3.411$ & $3.415$ &$3.330$& $3.459$\\
$2^4P_{\frac{5}{2}}$ & $0.046$ & $0.001$ & $-0.0002$ & $3.260 \pm 0.069 $ & $-$ &$3.435$ & $3.427$ &$3.290$& $3.471$\\
\hline
$3^2P_{\frac{1}{2}}$ & $-0.029$ & $-0.001$ & $-0.0005$ & $3.338 \pm 0.077$ & $-$ &$3.562$& $3.754$ &$3.644$& $3.62$ \\
$3^2P_{\frac{3}{2}}$ & $0.018$ & $-0.0002$ & $0.000$ & $3.386 \pm 0.076$ & $-$ &$3.566$& $3.752$ &$3.620$& $3.644$ \\
$3^4P_{\frac{1}{2}}$ & $-0.023$ & $-0.001$ & $-0.001$ & $3.343 \pm 0.077$ & $-$ &$3.554$&$3.717$ &$3.656$& $3.632$ \\
$3^4P_{\frac{3}{2}}$ & $-0.030$  & $-0.001$ & $0.0004$ &$3.339 \pm 0.077$& $-$ & $3.550$ & $3.737$ &$3.632$& $3.656$\\
$3^4P_{\frac{5}{2}}$ & $0.041$ & $0.0005$ & $-0.0001$ & $3.410 \pm 0.076$ & $-$ &$3.569$& $3.744$ &$3.601$& $3.601$\\
\hline
$4^2P_{\frac{1}{2}}$ & $-0.027$ & $-0.001$ & $-0.0003$ & $3.461 \pm0.082$ & $-$ &$3.735$& $4.037$ & $3.926$& $-$ \\
$4^2P_{\frac{3}{2}}$ & $0.016$ & $-0.0001$ & $0.000$ & $3.505 \pm 0.082$ & $-$ &$3.739$& $4.036$ & $3.903$& $-$ \\
$4^4P_{\frac{1}{2}}$ & $-0.021$ & $-0.001$ & $-0.0005$ & $3.465 \pm 0.082$ & $-$ &$3.728$& $4.009$ & $3.938$ & $-$\\
$4^4P_{\frac{3}{2}}$ & $-0.027$ & $-0.0004$ & $0.0002$ & $3.461 \pm 0.082$ & $-$ &$3.725$& $4.023$ & $3.915$ & $-$\\
$4^4P_{\frac{5}{2}}$ & $0.038$ & $0.0004$ & $-0.000$ & $3.527 \pm 0.082$ & $-$ &$3.742$& $4.028$ & $3.884$ & $-$\\
\hline
\end{tabular} 
\endgroup}
\end{table*}

\begin{table*}[ht]
{
\begingroup
\caption{$D$ State masses $\Omega_c^0$ (in $GeV$)} \label{tab:5}
\setlength{\tabcolsep}{5pt}
\renewcommand{\arraystretch}{1.5}
\begin{tabular}{ c c c c c c c c c c c}
\hline
\hline
$n^{2S+1}L_J$ & $\big<V^{jj}_{q_1q_2q_3}\big>$& $\big<V^{L.S}_{q_1q_2q_3}\big>$ & $\big<V^{T}_{q_1q_2q_3}\big>$& Our &Exp.\cite{ParticleDataGroup:2024cfk}&\cite{Yu:2022ymb}& \cite{Ebert:2011kk} & \cite{Shah:2016nxi}  & \cite{Jakhad:2023mni}\\
\hline
$1^2D_{\frac{3}{2}}$ & $-0.072$ & $-0.002$ & $-0.0002$ & $3.073 \pm 0.067$ &$3.065\pm0.21$&$3.313$& $3.298$ & $3.231$  & $3.323$\\
$1^2D_{\frac{5}{2}}$ & $-0.011$ & $-0.000$ & $0.000$ & $3.136 \pm 0.067$ &$-$&$3.314$& $3.297$ & $3.188$  &$3.364$\\
$1^4D_{\frac{1}{2}}$ & $-0.029$ & $-0.004$ & $-0.001$ & $3.113 \pm 0.067$ &$-$&$3.304$& $3.287$ & $3.215$  &$3.290$\\
$1^4D_{\frac{3}{2}}$ & $-0.027$ & $-0.003$ & $-0.0004$ & $3.118 \pm 0.067$ &$3.118\pm0.0001$&$3.304$& $3.282$ & $3.262$ &$3.334$\\
$1^4D_{\frac{5}{2}}$ & $0.054$ & $-0.001$ & $0.000$ & $3.201 \pm 0.066$ &$-$&$3.304$& $3.286$ & $3.173$  &$3.396$\\
$1^4D_{\frac{7}{2}}$ & $0.085$ & $0.002$ & $-0.000$ & $3.234 \pm 0.066$ &$-$&$3.315$& $3.283$ & $3.136$  &$3.423$\\
\hline 
$2^2D_{\frac{3}{2}}$ & $-0.059$ & $-0.001$ & $-0.000$ & $3.258 \pm 0.075$ &$-$&$3.650$& $3.627$ & $3.538$ & $-$\\
$2^2D_{\frac{5}{2}}$ & $-0.008$ & $-0.000$ & $0.000$ & $3.310 \pm 0.074$ &$-$&$3.651$& $3.626$ & $3.502$ &$-$\\
$2^4D_{\frac{1}{2}}$ & $-0.025$ & $-0.003$ & $-0.001$ & $3.290 \pm 0.075$ &$-$&$3.641$& $3.623$ & $3.524$ &$-$\\
$2^4D_{\frac{3}{2}}$ & $-0.023$ & $-0.002$ & $-0.000$ & $3.294 \pm 0.077$ &$-$&$3.641$& $3.613$ & $3.565$ &$-$\\
$2^4D_{\frac{5}{2}}$ & $0.045$ & $-0.000$ & $0.000$ & $3.362 \pm 0.074$ &$-$&$3.640$& $3.614$ & $3.488$ &$-$\\
$2^4D_{\frac{7}{2}}$ & $0.070$ & $0.001$ & $-0.000$ & $3.390 \pm 0.074$ &$-$&$3.652$& $3.611$ & $3.456$ &$-$\\
\hline
$3^2D_{\frac{3}{2}}$ & $-0.052$ & $-0.001$ & $-0.000$ & $3.394 \pm 0.080$ &$-$&$3.771$& $-$  &$-$&$-$\\
$3^2D_{\frac{5}{2}}$ & $-0.007$ & $-0.000$ & $0.000$ & $3.441 \pm 0.080$ &$-$&$3.772$& $-$ &$-$&$-$\\
$3^4D_{\frac{1}{2}}$ & $-0.022$ & $-0.002$ & $-0.000$ & $3.422 \pm 0.080$ &$-$&$3.764$& $-$  & $-$&$-$\\
$3^4D_{\frac{3}{2}}$ & $-0.021$ & $-0.001$ & $-0.000$ & $3.425 \pm 0.080$ &$-$&$3.764$& $-$  & $-$&$-$\\
$3^4D_{\frac{5}{2}}$ & $0.040$ & $-0.000$ & $0.000$ & $3.487 \pm 0.080$ &$-$&$3.764$& $-$ &$-$&$-$\\
$3^4D_{\frac{7}{2}}$ & $0.063$ & $0.001$ & $-0.000$ & $3.511 \pm 0.079$ &$-$&$3.773$& $-$  &$-$&$-$\\
\hline
$4^2D_{\frac{3}{2}}$ & $-0.048$ & $-0.001$ & $-0.000$ & $3.502 \pm 0.085$ &$-$&$3.917$& $-$ & $-$ & $-$\\
$4^2D_{\frac{5}{2}}$ & $-0.006$ & $-0.000$ & $0.000$ & $3.545 \pm 0.085$ &$-$&$3.918$& $-$ & $-$ & $-$\\
$4^4D_{\frac{1}{2}}$ & $-0.021$ & $-0.001$ & $-0.0002$ & $3.528 \pm 0.085$ &$-$&$3.909$& $-$ & $-$ & $-$\\
$4^4D_{\frac{3}{2}}$ & $-0.019$ & $-0.001$ & $-0.000$ & $3.531 \pm 0.085$ &$-$&$3.909$& $-$ & $-$ & $-$\\
$4^4D_{\frac{5}{2}}$ & $0.037$ & $-0.000$ & $0.000$ & $3.588 \pm 0.085$ &$-$&$3.908$& $-$ & $-$ & $-$\\
$4^4D_{\frac{7}{2}}$ & $0.058$ & $0.000$ & $-0.000$ & $3.610 \pm 0.084$ &$-$&$3.919$& $-$ & $-$ & $-$\\
\hline
\end{tabular}
\endgroup}
\end{table*}

\begin{table*}[ht]
{
\begingroup
\caption{$S$ State masses of $\Omega_b^{-}$(in $GeV$)} \label{tab:6}
\setlength{\tabcolsep}{5pt}
\renewcommand{\arraystretch}{1.5}
\begin{tabular}{c c c c c c c c c c }
\hline
\hline
$nL$ & $J^{P}$ & State & $\big<V^{jj}_{q_1q_2q_3}\big>$ & Our& Exp.\cite{ParticleDataGroup:2024cfk} & \cite{Jakhad:2024fin}& \cite{Kakadiya:2022zvy} &  \cite{Ebert:2011kk}  & \cite{Yu:2022ymb}\\
\hline
$1S$ & $\frac{1}{2}^+$ & $1^2S_{\frac{1}{2}}$ & $-0.023$ & $6.047\pm 0.066 $ & $6.045\pm0.0008$ & $6.065$ & $6.046$ & $6.064$ & $6.043$ \\
$1S$ & $\frac{3}{2}^+$ & $1^4S_{\frac{3}{2}}$ & $0.014$ & $6.083\pm 0.068$ &$-$& $6.081$& $6.082$ & $6.088$ & $6.069$  \\

$2S$ & $\frac{1}{2}^+$ & $2^2S_{\frac{1}{2}}$ & $-0.018$ & $6.436\pm 0.085$ & $-$ &$6.492$& $6.438$ & $6.450$ & $6.446$ \\
$2S$ & $\frac{3}{2}^+$ & $2^4S_{\frac{3}{2}}$ & $0.011$ & $6.465\pm 0.084$ & $-$ &$6.497$& $6.462$ & $6.461$ & $6.466$ \\

$3S$ & $\frac{1}{2}^+$ & $3^2S_{\frac{1}{2}}$ & $-0.016$ & $6.661\pm 0.095 $ & $-$ &$6.840$ & $6.740$ & $6.804$ & $6.633$\\
$3S$ & $\frac{3}{2}^+$ & $3^4S_{\frac{3}{2}}$ & $0.010$ & $6.687\pm 0.095$ & $-$ &$6.841$& $6.753$ & $6.811$ & $6.650$ \\

$4S$ & $\frac{1}{2}^+$ & $4^2S_{\frac{1}{2}}$ & $-0.015$ & $6.823\pm 0.102$ & $-$ &$7.141$& $7.022$ & $7.091$ & $6.790$ \\
$4S$ & $\frac{3}{2}^+$ & $4^4S_{\frac{3}{2}}$ & $0.009$ & $6.846\pm 0.102 $ & $-$ &$7.142$& $7.030$ & $7.096$ & $6.804$ \\

$5S$ & $\frac{1}{2}^+$ & $5^2S_{\frac{1}{2}}$ & $-0.014$ & $6.949\pm 0.108 $ & $-$ &$7.411$& $7.290$ & $7.338$ & $-$ \\
$5S$ & $\frac{3}{2}^+$ & $5^4S_{\frac{3}{2}}$ & $0.008$ & $6.971\pm 0.108 $ & $-$ &$7.412$& $7.296$  & $7.343$ & $-$\\
\hline
\end{tabular}
\endgroup}
\end{table*}

\begin{table*}[ht]
{
\begingroup
\caption{$P$ State masses of $\Omega^-_b$(in $GeV$)} \label{tab:7}
\setlength{\tabcolsep}{5pt}
\renewcommand{\arraystretch}{1.5}
\begin{tabular}{ c c c c c c c c c}
\hline
\hline
$n^{2S+1}L_J$ & $\big<V^{jj}_{q_1q_2q_3}\big>$& $\big<V_{q_1q_2q_3}^{L.S}\big>$ & $\big<V_{q_1q_2q_3}^T\big>$& Our & \cite{Jakhad:2024fin}&\cite{Kakadiya:2022zvy}& \cite{Ebert:2011kk} &\cite{Yu:2022ymb}\\
\hline
$1^2P_{\frac{1}{2}}$ & $-0.021$ & $-0.002$ & $-0.001$ & $6.323 \pm 0.080$ &$6.322$ & $6.344$ & $6.339$ &$6.334$\\
$1^2P_{\frac{3}{2}}$ & $0.013$ & $-0.000$ & $0.000$ & $6.360 \pm 0.080$ &$6.372$& $6.341$ & $6.340$ &$6.336$\\
$1^4P_{\frac{1}{2}}$ & $-0.017$ & $-0.003$ & $-0.002$ & $6.325 \pm 0.080$ &$6.350$&$6.345$ & $6.330$ & $6.329$\\
$1^4P_{\frac{3}{2}}$ & $-0.023$ & $-0.001$ & $0.001$ & $6.324 \pm 0.080$ &$6.443$& $6.343$ & $6.331$ & $6.326$\\
$1^4P_{\frac{5}{2}}$ & $0.031$ & $0.001$ & $-0.000$ &$6.379 \pm 0.079$ &$6.456$& $6.339$  & $6.334$ & $6.339$\\
\hline
$2^2P_{\frac{1}{2}}$ & $-0.017 $ & $-0.001$ & $-0.001$ & $6.584 \pm0.092$&$6.730$ &$6.596$ & $6.710$ &$6.662$ \\
$2^2P_{\frac{3}{2}}$ & $0.011$ & $-0.0002$ & $0.000$ & $6.614 \pm 0.091$ &$6.751$& $6.594$ & $6.705$ &$6.664$\\
$2^4P_{\frac{1}{2}}$ & $-0.014$ & $-0.001$ & $-0.001$ & $6.586 \pm 0.092$&$6.744$ &$6.597$ & $6.706$ &$6.658$\\
$2^4P_{\frac{3}{2}}$ & $-0.018$ & $-0.001$ & $0.000$ & $6.585 \pm 0.091$ &$6.783$& $6.595$ & $6.699$ & $6.655$\\
$2^4P_{\frac{5}{2}}$ & $0.025$ & $0.001$ & $-0.000$ & $6.628 \pm 0.091$ &$6.787$&$6.592$  & $6.700$ & $6.666$\\
\hline
$3^2P_{\frac{1}{2}}$ & $-0.015$ & $-0.001$ & $-0.000$ & $6.763 \pm 0.010$ &$7.051$&$6.829$ & $7.009$ & $6.844$ \\
$3^2P_{\frac{3}{2}}$ & $0.009$ & $-0.000$ & $0.000$ & $6.789 \pm 0.099$&$7.065$ &$6.827$ & $7.002$&$6.846$ \\
$3^4P_{\frac{1}{2}}$ & $-0.012$ & $-0.001$ & $-0.001$ & $6.766 \pm 0.010$&$7.061$ &$6.830$ & $7.003$ &$6.841$ \\
$3^4P_{\frac{3}{2}}$ & $-0.016$  & $-0.000$ & $0.000$ &$6.764 \pm 0.010$ &$7.087$& $6.828$  & $6.998$ &$6.839$\\
$3^4P_{\frac{5}{2}}$ & $0.022$ & $0.000$ & $-0.000$ & $6.802 \pm 0.099$ &$7.089$&$6.826$ & $6.996$ &$6.848$\\
\hline
$4^2P_{\frac{1}{2}}$ & $-0.014$ & $-0.0005$ & $-0.0002$ & $6.901 \pm 0.105$ &$7.333$ &$7.044$ & $7.265$ &$6.969$ \\
$4^2P_{\frac{3}{2}}$ & $0.009$ & $-0.000$ & $0.000$ & $6.924 \pm 0.105$ &$7.343$&$7.043$ & $7.258$ &$6.970$ \\
$4^4P_{\frac{1}{2}}$ & $-0.011$ & $-0.001$ & $-0.000$ & $6.903 \pm 0.106$&$7.341$ &$7.043$ & $7.257$ &$6.966$ \\
$4^4P_{\frac{3}{2}}$ & $-0.014$ & $-0.0003$ & $0.000$ & $6.901 \pm 0.106$ &$7.361$&$7.043$ & $7.250$ &$6.964$ \\
$4^4P_{\frac{5}{2}}$ & $0.020$ & $0.0002$ & $-0.000$ & $6.936 \pm 0.105$ &$7.363$&$7.042$ & $7.251$ &$6.972$ \\
\hline
\end{tabular}
\endgroup}
\end{table*}

\begin{table*}[ht]
{
\begingroup
\caption{$D$ State masses of $\Omega_b^-$ (in $GeV$)} \label{tab:8}
\setlength{\tabcolsep}{5pt}
\renewcommand{\arraystretch}{1.5}
\begin{tabular}{ c c c c c c c c c }
\hline
\hline
$n^{2S+1}L_J$ & $\big<V^{jj}_{q_1q_2q_3}\big>$& $\big<V^{L.S}_{q_1q_2q_3}\big>$ & $\big<V^{T}_{q_1q_2q_3}\big>$& Our &\cite{Jakhad:2024fin}& \cite{Ebert:2011kk} & \cite{Kakadiya:2022zvy} & \cite{Yu:2022ymb}\\
\hline
$1^2D_{\frac{3}{2}}$ & $-0.039$ & $-0.002$ & $-0.000$ & $6.488 \pm 0.089$ &$6.631$& $6.549$ & $6.480$ & $6.561$\\
$1^2D_{\frac{5}{2}}$ & $-0.006$ & $-0.000$ & $0.000$ & $6.523 \pm 0.088$ &$6.727$& $6.529$ & $6.476$ & $6.561$\\
$1^4D_{\frac{1}{2}}$ & $-0.016$ & $-0.003$ & $-0.001$ & $6.509 \pm 0.089$ &$6.620$& $6.540$ & $6.485$ & $6.556$\\
$1^4D_{\frac{3}{2}}$ & $-0.014$ & $-0.002$ & $-0.000$ & $6.512 \pm 0.089$ &$6.663$& $6.530$ & $6.482$ & $6.556$\\
$1^4D_{\frac{5}{2}}$ & $0.029$ & $-0.001$ & $0.000$ & $6.558 \pm 0.089$ &$6.673$& $6.520$ & $6.478$ & $6.555$\\
$1^4D_{\frac{7}{2}}$ & $0.046$ & $0.001$ & $-0.000$ & $6.576 \pm 0.087$ &$6.736$& $6.517$ & $6.472$ & $6.562$\\
\hline
$2^2D_{\frac{3}{2}}$ & $-0.031$ & $-0.001$ & $-0.000$ & $6.691 \pm 0.097$ &$6.971$& $6.863$ & $6.726$ & $6.852$\\
$2^2D_{\frac{5}{2}}$ & $-0.004$ & $-0.000$ & $0.000$ & $6.718 \pm 0.097$ &$7.033$& $6.846$ & $6.723$ & $6.850$\\
$2^4D_{\frac{1}{2}}$ & $-0.013$ & $-0.002$ & $-0.000$ & $6.707 \pm 0.097$ &$6.965$& $6.857$ & $6.730$ & $6.846$\\
$2^4D_{\frac{3}{2}}$ & $-0.012$ & $-0.001$ & $-0.000$ & $6.709 \pm 0.097$ &$6.992$& $6.846$ & $6.727$ & $6.846$\\
$2^4D_{\frac{5}{2}}$ & $0.024$ & $-0.000$ & $0.000$ & $6.746 \pm 0.096$ &$6.998$& $6.837$ & $6.724$ & $6.846$\\
$2^4D_{\frac{7}{2}}$ & $0.037$ & $0.001$ & $-0.000$ & $6.761 \pm 0.096$ &$7.038$& $6.834$ & $6.720$ & $6.853$\\
\hline
$3^2D_{\frac{3}{2}}$ & $-0.028$ & $-0.001$ & $-0.000$ & $6.841 \pm 0.104$ &$7.262$& $-$ & $6.953$ & $7.026$\\
$3^2D_{\frac{5}{2}}$ & $-0.003$ & $-0.000$ & $0.000$ & $6.865 \pm 0.103$ &$7.311$& $-$ & $6.951$ & $7.026$\\
$3^4D_{\frac{1}{2}}$ & $-0.012$ & $-0.001$ & $-0.000$ & $6.856 \pm 0.104$ &$7.258$& $-$ & $6.956$ & $7.021$\\
$3^4D_{\frac{3}{2}}$ & $-0.011$ & $-0.001$ & $-0.000$ & $6.857 \pm 0.104$ &$7.280$& $-$ & $6.954$ & $7.022$\\
$3^4D_{\frac{5}{2}}$ & $0.021$ & $-0.000$ & $0.000$ & $6.890 \pm 0.103$ &$7.283$& $-$ & $6.951$ & $7.021$\\
$3^4D_{\frac{7}{2}}$ & $0.033$ & $0.001$ & $-0.000$ & $6.903 \pm 0.103$ &$7.313$& $-$ & $6.948$ & $7.027$\\
\hline
$4^2D_{\frac{3}{2}}$ & $-0.026$ & $-0.001$ & $-0.000$ & $6.961 \pm 0.109$ &$-$& $-$ & $7.164$ & $7.124$\\
$4^2D_{\frac{5}{2}}$ & $-0.003$ & $-0.000$ & $0.000$ & $6.984 \pm 0.109$ &$-$& $-$ & $7.162$ & $7.124$\\
$4^4D_{\frac{1}{2}}$ & $-0.011$ & $-0.001$ & $-0.000$ & $6.975 \pm 0.109$ &$-$& $-$ & $7.166$ & $7.121$\\
$4^4D_{\frac{3}{2}}$ & $-0.010$ & $-0.001$ & $-0.000$ & $6.976 \pm 0.109$ &$-$& $-$ & $7.164$ & $7.121$\\
$4^4D_{\frac{5}{2}}$ & $0.020$ & $-0.000$ & $0.000$ & $7.006 \pm 0.109$ &$-$& $-$ & $7.162$ & $7.121$\\
$4^4D_{\frac{7}{2}}$ & $0.031$ & $0.000$ & $-0.000$ & $7.018 \pm 0.108$ &$-$& $-$ & $7.160$ & $7.125$\\
\hline
\end{tabular}
\endgroup}
\end{table*}

\section{Magnetic moments and Decay properties}
\label{sec:3}
The magnetic moment of baryons is expressed in relation to its constituent quarks \cite{Patel:2007gx} as follows:
\begin{eqnarray}
    \mu_{B} = \sum_{q}\bigg<\phi_{sf}|\vec{\mu}_{qz}|\phi_{sf}\bigg>, 
    \end{eqnarray}
    where
    \begin{eqnarray}\label{m}
    \mu_{q} = \frac{e_{q}}{2 m_{q}}\sigma_{q}. 
    \end{eqnarray}
Here, $e_q$ and $\sigma_q$ represent the charge and the spin of the quark, and $|\phi_{sf}\big>$ is the spin-flavor wave function. The spin-flavor wave functions of $\Omega_c^0$ \& $\Omega_b^-$ baryons with $J^P=1/2^+$ are
\begin{multline}
    |\phi_{sf}\big>_{\Omega_c^0} = \frac{\sqrt{2}}{6}(2c_- s_+s_+ - s_-c_+s_+ - c_+s_-s_+ \\+ 2s_+c_-s_+ - s_+s_-c_+- s_-s_+c_+\\-s_+c_+s_--c_+s_+s_-+2s_+s_+c_-),
\end{multline}
\begin{multline}
    |\phi_{sf}\big>_{\Omega_b^-} = \frac{\sqrt{2}}{6}(2b_- s_+s_+ - s_-b_+s_+ - b_+s_-s_+ \\+ 2s_+b_-s_+ - s_+s_-b_+- s_-s_+b_+\\-s_+b_+s_--b_+s_+s_-+2s_+s_+b_-),
\end{multline}
and the spin-flavor wave functions of $\Omega_c^0$ \& $\Omega_b^-$ baryons with $J^P=3/2^+$ are
\begin{multline}
    |\phi_{sf}\big>_{\Omega_c^0} = \frac{1}{\sqrt{3}}(s_+ s_+ c_+ + s_+ c_+ s_+ + c_+ s_+ s_+),
\end{multline}
\begin{multline}
    |\phi_{sf}\big>_{\Omega_b^-} = \frac{1}{\sqrt{3}}(s_+ s_+ b_+ + s_+ b_+ s_+ + b_+ s_+ s_+).
\end{multline}
Here we have used the notations for spin operator $\hat{S}|q\uparrow\big>=\hat{S}|q+\big>$ \& $\hat{S}|q\downarrow\big>=\hat{S}|q-\big>$.
Due to the interactions of three quarks inside the bound system like a baryon, the masses of the quarks may vary according to the interaction. We refer to them as the effective quark mass. The effective quark masses $m_{q}^{eff}$ in our model is defined as
\begin{eqnarray} 
    m_{q}^{eff} = E^D_q\left(1+\frac{\big<H\big>-E_{CM}}{\sum_q E^D_q}\right).
\end{eqnarray}
Here, the $\big<H\big>$ includes the strength of spin-spin interactions only as we are calculating the magnetic field of $S$ waves.
The effective quark mass follows the property of $M_J = \sum_{q=1}^{3} m_{q}^{eff}$. Our prediction and comparison with other different approaches are given in Table \ref{tab:9}.

\begin{table}[ht]   
{\begingroup
\caption{Magnetic moments {(in $\mu_N$)}} \label{tab:9}
\setlength{\tabcolsep}{5pt}
\renewcommand{\arraystretch}{1.5}
\begin{tabular}{ c c c c c c }
\hline
\hline
State & Our & \cite{Bernotas:2012nz}& \cite{Patel:2007gx} & \cite{Kakadiya:2022zvy} &\cite{Shah:2016nxi}\\
\hline
$\Omega^{0}_c \frac{1}{2}^+$ & $-0.860$ & $–0.774$& $-0.916$ & $-$&$-0.842$ \\
$\Omega^{0}_c \frac{3}{2}^+$ & $-0.667$ & $-0.547$& $-0.827$ & $-$ & $-0.625$ \\
$\Omega^{-}_b \frac{1}{2}^+$ & $-0.539$& $–0.545$ & $-$ & $-0.761$ &$-$\\
$\Omega^{-}_b \frac{3}{2}^+$ & $-0.906$ & $–0.919$& $-$ & $-1.236$ &$-$ \\
\hline
\end{tabular}
\endgroup}
\end{table}

\subsection{Radiative decay}\label{sec:3.1}
Radiative decays of baryons provide valuable insights into their intrinsic structure and the relationship between the decay process and the masses of their constituent quarks. The radiative decay width of singly heavy baryons, such as $\Omega_c^0$ \& $\Omega_b^-$, is comparatively larger than that of light baryons, hence it is significant enough to warrant detailed investigation. 

The electromagnetic radiative decay width can be expressed in terms of the radiative transition magnetic moment (in $ \mu_N $) and the photon energy ($ q = M_{3/2} - M_{1/2} $) as follows \cite{Dey:1994qi, Thakkar:2010ij}:  
\begin{eqnarray}
\Gamma_R = \frac{q^3}{4\pi} \frac{2}{2J+1} \frac{e^2}{m_p^2} |\mu_{\frac{3}{2}^+ \rightarrow \frac{1}{2}^+}|^2,
\end{eqnarray}  
where the transition magnetic moment is given by:  
\begin{multline}
\mu_{\frac{3}{2}^+ \rightarrow \frac{1}{2}^+} = \sum_i \big< \phi_{sf}^{\frac{3}{2}^+} | \mu_i \cdot \vec{\sigma_i} | \phi_{sf}^{\frac{1}{2}^+} \big> \\  
= \frac{2\sqrt{2}}{3} (\mu_u - \mu_s).
\end{multline}  

A key aspect of this calculation is the determination of the quark magnetic moments, which are obtained by taking the geometric mean of their effective masses. This is expressed as \cite{Thakkar:2010ij, Dhir:2009ax}:  
\begin{eqnarray}
m_i^{eff} = \sqrt{m_{i(\frac{3}{2}^+)}^{eff} \, m_{i(\frac{1}{2}^+)}^{eff}}.
\end{eqnarray}  
Our calculated transition magnetic moments and decay widths are mentioned in the table \ref{tab:r} and  calculated decay widths from other approaches are mentioned in the table for comparison. 

\begin{table}[ht]
{\begingroup
\centering
\caption{Transition magnetic moment(TMM)(in $\mu_N$) and Decay Width(DW){(in keV)}} \label{tab:r}
\setlength{\tabcolsep}{5pt}
\renewcommand{\arraystretch}{1.5}
\begin{tabular}{ c c c c c  }
\hline
\hline
Baryon & TMM & DW & \cite{Shah:2016nxi} & \cite{Kakadiya:2022zvy}  \\
\hline
$\Omega^{0}_c(1S^4\frac{3}{2}^+ \to 1S^2\frac{1}{2}^+)$ & $-0.893674$ & $1.172$ & $1.441$& $-$ \\
$\Omega^{-}_b(1S^4\frac{3}{2}^+ \to 1S^2\frac{1}{2}^+)$ & $-0.166325$ & $0.0057$ & $-$ & $0.01$\\
\hline
\end{tabular}
\endgroup}
\end{table}
\subsection{Non-leptonic Weak decays of $\Omega_c^0$}\label{sec:3.2}
The final products of non-leptonic weak decays of $\Omega_c^0$ involve octet ($B$) or decuplet ($B^*$) baryon along with pseudoscalar ($M$) or vector meson(V). To establish the strict $SU(3)_f$ relations for possible decays and to parameterize the $W$- exchange processes and $W$- emission, the topological diagram approach is used in \cite{Hsiao:2023mvw}. These two body decays of $\Omega_c^0$ involve, $c \to su\bar{d}$ \& $c \to qu\bar{q}$ weak quark-level transitions, where $q=(d,s)$. The effective Hamiltonian is
\begin{eqnarray}
    \mathcal{H}_{eff} = \sum_{i=+,-}\frac{G_F}{\sqrt{2}}c_i(V_{cs}^* V_{ud}O_i+V_{cq}^* V_{uq}O_i^q).
\end{eqnarray}
Where $O$'s represent the four quark operators, 
\begin{multline}
    O_{\pm} = \frac{1}{2}[(\bar{u}d)(\bar{s}c)\pm(\bar{s}d)(\bar{u}c)],\\
    O_{\pm}^q = \frac{1}{2}[(\bar{u}q)(\bar{q}c)\pm(\bar{q}q)(\bar{u}c)],
\end{multline}
with $(\bar{q}_aq_b)\equiv\bar{q}_a\gamma_\mu(1-\gamma_5)q_b$. The irreducible $SU(3)_f$ approach is used to aid the derivation of approximate relations for the topological parameters. The amplitudes $\mathcal{M}_{STDA}$ associated with this effective Hamiltonian are parameterized; their parametric form is given in the TABLE \ref{tab:10}, and all the numeric values of the parameters are given in \cite{Hsiao:2023mvw}. The branching fraction in terms of these parametrized amplitudes is given by,
\begin{multline} \label{branching fraction}
    B(\Omega^0_c \rightarrow B^{(*)} M,B^{(*)}V)=\frac{G^2_{F} |\vec{p}_{cm}| \tau_{\Omega_c}}{16 \pi m^2_{\Omega_c}} \\|\mathcal{M}_{STDA}(\Omega^0_c \rightarrow  B^{(*)} M,B^{(*)}V)|^2,
\end{multline} 
where the magnitude of the three-momentum of the final state is,  
\begin{eqnarray}
  |\vec{p}_{cm}| = \frac{\left[(m^2_{\Omega_c} - m^2_+) (m^2_{\Omega_c}-m^2_-)\right]^\frac{1}{2}}{2m_{\Omega_c}},     
\end{eqnarray}
with $m{\pm} =m_B^* \pm m_M(V)$, $\tau_{\Omega_c}$ is the lifetime of $\Omega^0_c $ baryon. 
Our predictions for the non-leptonic weak decay branching ratios and relative branching ratio of $ \Omega_c^0 $, along with the corresponding experimental values and theoretical predictions from alternative approaches, are presented in Table \ref{tab:10}.

\begin{table*}[ht]
{\begingroup
\caption{Non-leptonic weak decay branching ratio (BR) and relative branching ratios (RBR) for $\Omega_c^0$ decays along with corresponding $\mathcal{M}_{STDA}$} \label{tab:10}
\setlength{\tabcolsep}{5pt}
\renewcommand{\arraystretch}{1.5}
\begin{tabular}{ l c c c c c }
\hline
\hline
Decay & $\mathcal{M}_{STDA}$ & Our BR $(10^{-4})$ & \cite{Hsiao:2023mvw} & Our RBR & Exp. RBR\cite{ParticleDataGroup:2024cfk}\\
\hline
$\Omega^{0}_c \rightarrow \Xi^0 \bar{K^0}$ & $-(c+c')$ & $52.74$ & $47 \pm 7.2$ & $1.62$ & $1.64 \pm 0.29$\\
$\Omega^{0}_c \rightarrow \Lambda^0 \bar{K^0}$ & $\frac{2}{\sqrt{6}}c's_c$ & $10.60$ & $9 \pm0.4 $ & $0.33$ & $-$
\\
$\Omega^{0}_c \rightarrow \Xi^- \pi{^+}$ & $-ts_c$ & $49.74$ & $4.7 \pm 0.4 $ & $0.15$ & $0.25\pm 0.06$
\\
$\Omega^{0}_c \rightarrow \Xi^0 \pi{^0}$ & $-\frac{1}{\sqrt{2}}cs_c$ & $2.48$ & $2.3\pm 0.2 $ & $0.08$ & $-$
\\
$\Omega^{0}_c \rightarrow \Xi^0 \eta{^0}$ & $\frac{1}{\sqrt{6}}(3c+2c')s_c$ & $0.20$ & $0.2\pm 0.1 $ & $0.006$ & $-$
\\
$\Omega^{0}_c \rightarrow \Xi^- \rho{^+}$ & $-\bar{t}s_c$ & $17.55$ & $16.4\pm 1.3 $ & $0.54$ & $-$
\\
$\Omega^{0}_c \rightarrow \Xi^{*0} \bar{K^0}$ & $\frac{1}{\sqrt{3}}c_d$ & $10.72$ & $9.8 \pm 1.3$ & $0.33$ & $-$
\\
$\Omega^{0}_c \rightarrow \Omega^{-} \pi^+$ & $t_d$ & $32.47$ & $28.1 \pm 3.6$ & $1$ & $1.8$
\\
$\Omega^{0}_c \rightarrow \Omega^{-} K^+$ & $(t_d-e_{dM}+\delta e^s_{dB})s_c$ & $2.03$ & $1.8 \pm 0.3$ & $0.06$ & $<0.29$
\\
$\Omega^{0}_c \rightarrow \Sigma^{*+} K^-$ & $\frac{1}{3}e_{dM}s_c$ & $2.92$ & $2.7 \pm 0.2$ & $0.09$ & $-$
\\
$\Omega^{0}_c \rightarrow \Xi^{*-} \pi^+$ & $-\frac{1}{\sqrt{3}}(t_d+e_{dM})s_c$ & $6.12$ & $5.9\pm 0.3$ & $0.19$ & $-$
\\
$\Omega^{0}_c \rightarrow \Xi^{*0} \pi^0$ & $\frac{1}{\sqrt{6}}(c_d-e_{dM})s_c$ & $3.05$ & $2.9\pm 0.2$ & $0.09$ & $-$
\\
$\Omega^{0}_c \rightarrow \Xi^{*0} \eta^0$ & $-\frac{1}{\sqrt{2}}(c_d+e_{dM}+\frac{2}{3}\delta e^s_{dM})s_c$ & $1.12$ & $1.1\pm 0.2$ & $0.03$ & $-$
\\
$\Omega^{0}_c \rightarrow \Omega^{-} \rho^+$ & $\bar{t}_d$ & $55.80$ & $46.3\pm 1.72$ & $1.72$ & $>1.3$
\\
$\Omega^{0}_c \rightarrow \Omega^{-} K^{*+}$ & $(\bar{t}_d-\bar{e}_{dM}+\delta \bar{e}^s_{dB})s_c$ & $3.01$ & $2.5\pm 1.1$ & $0.09$ & $-$
\\
$\Omega^{0}_c \rightarrow \Sigma^{*+} K^{*-}$ & $\frac{1}{\sqrt{3}}\bar{e}_{dM}s_c$ & $6.03$ & $5.7\pm 1.3$ & $0.18$ & $-$
\\
$\Omega^{0}_c \rightarrow \Sigma^{*0} K^{*0}$ & $\frac{1}{\sqrt{6}}\bar{e}_{dM}s_c$ & $3.15$ & $2.8\pm 0.6$ & $0.1$ & $-$
\\
$\Omega^{0}_c \rightarrow \Xi^{*-} \rho^{*+}$ & $-\frac{1}{\sqrt{3}}(\bar{t}_d+\bar{e}_{dM})s_c$ & $11.99$ & $11.7\pm 1.8$ & $0.37$ & $-$\\
\hline

\end{tabular}
\endgroup}
\end{table*}

\section{Non leptonic decays of $\Omega_b$ baryons}
The color-allowed two-body nonleptonic decays of bottom baryons $\Omega_b$ with the emission of a pseudoscalar ($\pi^{-}$, $K^{-}$, $D^{-}$, and $D_s^{-}$) or a vector meson ($\rho^{-}$, $K^{*-}$, $D^{*-}$, and $D_s^{*-}$) within na\"{i}ve factorization approach. Under this approximation, the hadronic transition matrix element is factorized into a product of two independent matrix elements \cite{Ke:2019smy}. Accordingly, we can express, 
\begin{multline}
\langle\mathcal{B}_c^{(*)}(P^{\prime},J_{z}^{\prime}) \ M^{-} \ |\mathcal{H}_{\text{eff}}| \ \mathcal{B}_b(P,J_{z})\rangle\\
=\frac{G_F}{\sqrt{2}} V_{cb} V_{qq^{\prime}}^{*}
\langle M^{-}|\bar{q}^{\prime}\gamma_{\mu}(1-\gamma_{5})q|0\rangle\\
\times\langle\mathcal{B}_c^{(*)}(P^{\prime},J_{z}^{\prime})|\bar{c}\gamma^{\mu}(1-\gamma_{5})b|\mathcal{B}_b(P,J_{z})\rangle,
\end{multline}
where the meson transition term is given by
\begin{equation}
\langle M|\bar{q}^{\prime}\gamma_{\mu}(1-\gamma_{5})q|0\rangle
=\left\{
\begin{array}{ll}
if_{P}q_{\mu}, &M=P\\
if_{V}\epsilon_{\mu}^{*}m_{V}, &M=V
\end{array}.
\right.
\end{equation}
In the case where the final state includes a pseudoscalar meson, the decay width \cite{Ke:2019smy} takes the following form 
\begin{multline}
\Gamma=\frac{|p_c|}{8\pi}\Big(\frac{(M+M^{\prime})^2-m^2}{M^2}|A|^2\\+\frac{(M-M^{\prime})^2-m^2}{M^2}|B|^2\Big),
\label{eq:PhysicsP}
\end{multline}
\begin{equation}
    \alpha=\frac{2\kappa\text{Re}(A^*B)}{|A|^2+\kappa^2|B|^2},
\end{equation}
where as for the transitions emitting vector meson in final states \cite{Ke:2019smy} we can write 
\begin{multline}
\Gamma=\frac{|p_c|(E^{\prime}+M^{\prime})}{4\pi M}\Big(2(|S|^2+|P_2|^2)\\+\frac{E_m^2}{m^2}(|S+D|^2+|P_1|^2)\Big),
\label{eq:PhysicsV} 
\end{multline}
\begin{eqnarray}
    \alpha=\frac{4m^2\text{Re}(S^*P_2)+2E_m^2\text{Re}(S+D)^*P_1}{2m^2(|S|^2+|P_2|^2)+E_m^2(|S+D|^2+|P_1|^2)},
\end{eqnarray}
Here, $p_c$ represents the momentum of the daughter baryon measured in the rest frame of the parent baryon, and $\kappa = |p_c|/(E^{\prime}+M^{\prime})$. Additionally, $M$ ($E$) and $M'$ ($E'$) denote the masses (energies) of the parent and daughter baryons, respectively, while $m$ ($E_m$) corresponds to the mass (energy) of the final-state meson. The amplitudes $A$ and $B$ in Eqs. (\ref{eq:PhysicsP}) are given by
\begin{equation}
A=\frac{G_F}{\sqrt{2}}V_{cb}V_{qq^{\prime}}^*a_{i}f_{P}(M-M^{\prime})f_1^V(m^2),
\end{equation}
\begin{equation}
B=-\frac{G_F}{\sqrt{2}}V_{cb}V_{qq^{\prime}}^*a_{i}f_{P}(M+M^{\prime})g_1^A(m^2),
\end{equation}
and $S$, $P_{1,2}$ and $D$ in Eqs. (\ref{eq:PhysicsV}) are expressed as
\begin{equation}
S=A_1,    
\end{equation}
\begin{equation}
P_1=-\frac{|p_c|}{E_m}\left(\frac{M+M^{\prime}}{E^{\prime}+M^{\prime}}B_1+MB_2\right),    
\end{equation}
\begin{equation}
P_2=\frac{|p_c|}{E^{\prime}+M^{\prime}}B_1,    
\end{equation}
\begin{eqnarray}
D=\frac{|p_c|^2}{E_m(E^{\prime}+M^{\prime})}(A_1-MA_2)    
\end{eqnarray}
with
\begin{multline}
    A_{1}=\frac{G_F}{\sqrt{2}}V_{cb}V_{qq^{\prime}}^*a_{i}f_{V}m_{V}\Big(g_1^A(m^2)\\+g_2^A(m^2)\frac{M-M^{\prime}}{M}\Big),
\end{multline}
\begin{eqnarray}
    A_{2}=\frac{G_F}{\sqrt{2}}V_{cb}V_{qq^{\prime}}^*a_{i}f_{V}m_{V}\left(2g_{2}^A(m^2)\right),
\end{eqnarray}
\begin{multline}
B_{1}=\frac{G_F}{\sqrt{2}}V_{cb}V_{qq^{\prime}}^*a_{i}f_{V}m_{V}\Big(f_1^V(m^2)\\-f_2^V(m^2)\frac{M+M^{\prime}}{M}\Big)    
\end{multline}
\begin{equation}
B_{2}=\frac{G_F}{\sqrt{2}}V_{cb}V_{qq^{\prime}}^*a_{i}f_{V}m_{V}\left(2f_{2}^V(m^2)\right),
\end{equation}

\begin{table*}[htbp]\centering
\caption{The branching ratio for $\Omega_b^-\to\Omega_c^0M^-$. All the values are multiplied by a factor of $10^{-3}$.}
\label{tab:Comparison}
\renewcommand\arraystretch{1.5}
\begin{tabular*}{178mm}{l@{\extracolsep{\fill}}ccccccc}
\hline
\hline
& This Work &\cite{Li:2021kfb}     & \cite{Cheng:1996cs}    & \cite{Ivanov:1997hi,Ivanov:1997ra}    & \cite{Zhao:2018zcb}  & \cite{Gutsche:2018utw}  & \cite{Chua:2019yqh}  \\
\hline
$\Omega_b^-\to\Omega_c^0\pi^-$  &  2.54  & 2.82         &4.92        &5.81            &4.00           &1.88  &$1.10^{+0.85}_{-0.55}$                               \\
$\Omega_b^-\to\Omega_c^0K^-$  &   0.20   &0.22         &$-$        &$-$             &0.326          &$-$   &$0.08^{+0.07}_{-0.04}$                               \\
$\Omega_b^-\to\Omega_c^0D^-$   &  0.46   & 0.52         & $-$         &$- $        &0.636          &- &$0.15^{+0.14}_{-0.08}$                               \\
$\Omega_b^-\to\Omega_c^0D_s^-$   & 12.2  & 13.5         & 17.9        &$-$             &17.1           & $-$   &$4.03^{+3.72}_{-2.21}$                               \\
\hline \\ 
$\Omega_b^-\to\Omega_c^0 (2S)\pi^-$   & 0.27  &  0.30        &  $- $   &     $-$    &   $-$  & $-$& $-$                              \\
$\Omega_b^-\to\Omega_c^0 (2S) K^-$    & 0.02   &  0.02        &    $-$   &$-$   &   $-$  & $-$  & $-$                              \\
$\Omega_b^-\to\Omega_c^0 (2S) D^-$   &  0.03   &  0.03        &   $ -$     &        $-$    &      $- $  &  $-$ &   $-$                            \\
$\Omega_b^-\to\Omega_c^0 (2S) D_s^-$   & 0.53  &  0.62        &    $-$    &       $-$     &      $- $    &  $-$ & $-$  \\ \hline \\   
$\Omega_b^-\to\Omega_c^0\rho^-$  & 7.43  & 7.92         &12.8        &$-$             &10.8           &5.43  &$3.07^{+2.41}_{-1.53}$                               \\
$\Omega_b^-\to\Omega_c^0K^{*-}$   & 0.32 & 0.41         &$-$         &$-$             &0.544          &$-$   &$0.16^{+0.12}_{-0.08}$                               \\
$\Omega_b^-\to\Omega_c^0D^{*-}$   & 0.39  &0.48         &$-$         &$-$             &0.511          &$-$   &$0.16^{+0.13}_{-0.08}$                               \\
$\Omega_b^-\to\Omega_c^0D_s^{*-}$ & 9.44 &9.73         &11.5        &$-$             &11.7           &$-$   &$3.18^{+2.69}_{-1.61}$                               \\
\hline \\
$\Omega_b^-\to\Omega_c^0(2S)\rho^-$  & 0.63  & 0.70 &    $-$    &     $-$      &     $-$   & $-$ &  $-$                             \\
$\Omega_b^-\to\Omega_c^0(2S)K^{*-}$   & 0.03 &  0.03       &   $-$     &     $-$       &      $- $   &  $-$ & $-$                              \\
$\Omega_b^-\to\Omega_c^0(2S)D^{*-}$   & 0.02  &   0.02      &     $-$    &      $- $     &     $- $    & $-$& $-$                               \\

$\Omega_b^-\to\Omega_c^0(2S)D_s^{*-}$ & 0.22 &   0.36      &    $-$   &    $- $   &    $- $     &  $-$ &   $-$                            \\
\hline
\end{tabular*}
\end{table*}

For numerical evaluation, the values of the form factors we adopted from the three-body light front quark model from Ref. \cite{Li:2021kfb}. For the color allowed transition, we consider the coefficient $a_1 = 1.018$  \cite{Chua:2019yqh}. The lifetime is taken as 
$\tau_{\Omega_b^-} = 1.65~\text{ps}$ \cite{ParticleDataGroup:2024cfk}. The CKM matrix elements used as \cite{ParticleDataGroup:2024cfk} 
\begin{equation*}
\begin{split}
& V_{cb}=0.0405,\ V_{ud}=0.9740,\ V_{us}=0.2265,\\
& V_{cd}=0.2264,\ V_{cs}=0.9732.
\end{split}
\end{equation*}
The decay constants of pseudoscalar and vector mesons are taken from \cite{ParticleDataGroup:2024cfk}
\begin{equation*}
\begin{split}
&f_{\pi}=130.2,\ f_{K}=155.6,\ f_{D}=211.9,\ f_{D_s}=249.0,\\
&f_{\rho}=216,\ f_{K^*}=210,\ f_{D^*}=220,\ f_{D_s^*}=230.
\end{split}
\end{equation*}

In the present work, the branching ratios are predicted using the baryon masses calculated within our formalism. Table \ref{tab:Comparison} compares our results for the branching fractions of $\Omega_b^- \to \Omega_c^0M^-$ decays with those from various theoretical models, including the nonrelativistic quark model \cite{Cheng:1996cs}, the relativistic three-quark model \cite{Ivanov:1997hi,Ivanov:1997ra}, the light-front quark model \cite{Zhao:2018zcb,Chua:2019yqh}, and the covariant confined quark model \cite{Gutsche:2018utw}. This comparison highlights the consistency and differences between our predictions and previous studies.

\section{Discussion and Conclusion}\label{sec:4}
We have investigated the applicability of the modified independent quark model within a relativistic framework to predict the masses, magnetic moments, and decay properties of singly heavy baryons, specifically $ \Omega_c^0 $ and $ \Omega_b^- $. These baryons have been recently observed, yet their spin-parity assignments remain uncertain. Our predictions for the ground and first excited states of $ \Omega_c^0 $ show good agreement with experimental observations. Additionally, we have attempted to determine the spin-parity of five out of the seven observed states. 
We propose the spin-parity ($ J^P $) assignments of $ \Omega_c^0(3050) $, $ \Omega_c^0(3065) $, $ \Omega_c^0(3120)$, and $ \Omega_c^0(3185) $ as $ \frac{5}{2}^- $, $ \frac{3}{2}^+ $, $ \frac{3}{2}^+ $, and $ \frac{1}{2}^- $, respectively, based on our mass predictions presented in Tables \ref{tab:3}, \ref{tab:4}, and \ref{tab:5} . These results align with observational constraints on spin-parity assignments \cite{LHCb:2021ptx}; for instance, the $ J = \frac{1}{2} $ hypothesis is rejected at $ 2.2\sigma $ for $ \Omega_c^0(3050) $ and at $ 3.6\sigma $ for $ \Omega_c^0(3065) $. Due to the close mass spacing of the $ P $-wave states, the spin-parity of $ \Omega_c^0(3000) $ remains uncertain and could be either $ \frac{1}{2}^- $ or $ \frac{3}{2}^- $. 
Our predicted masses for the $ L = 1 $ orbital excitation states of $ \Omega_b^- $ (Table \ref{tab:7}) fall within the range of the four recently observed excited states of $ \Omega_b^- $(ie., $\Omega_b^-(6316)$, $\Omega_b^-(6330)$, $\Omega_b^-(6340)$, \& $\Omega_b^-(6350)$), in agreement with experimental results \cite{LHCb:2020tqd}. While we could not assign specific spin-parity values to these states, they are most likely $ P $-wave excitations of $ \Omega_b^- $. Additionally, our predictions for higher excited states may serve as a reference for future observations. The mass uncertainties presented in Tables \ref{tab:3}, \ref{tab:4}, \ref{tab:5}, \ref{tab:6}, \ref{tab:7}, and \ref{tab:8} arise solely from a \( 5\% \) variation in the fitted potential parameters listed in Tables \ref{tab:1} and \ref{tab:2}. These variations were introduced to assess the sensitivity of our predictions to changes in the model parameters.

We have calculated the magnetic moments of the ground and first radially excited states of $ \Omega_c^0 $ and $ \Omega_b^- $ using the effective masses of quarks. Our predictions fall within the range of those obtained from other theoretical approaches, as given in Table \ref{tab:9}. 

We have also calculated these two baryons' radiative decay width and electromagnetic transition magnetic moments and mentioned them in Table \ref{tab:r}. Despite the small decay widths, we observe that the decay width of $ \Omega_b^- $ is significantly smaller than that of $ \Omega_c^0 $. While these decays may never be experimentally observed, they offer valuable insights into the electromagnetic interactions within singly heavy baryons.
Finally, it is essential to study and calculate other decay properties of a baryon that can be experimentally verified to assess the validity of our model. To this end, we have computed the decay widths and branching ratios for various non-leptonic decays of $ \Omega_c^0 $. For simplification, we have normalized the branching ratio of $ \Omega_c^0 \rightarrow \Omega^- \pi^+ $ to 1 and determined the relative branching ratios of other decay modes accordingly. As shown in Table \ref{tab:10}, our predicted branching ratio for the decay $ \Omega_c^0 \to \Xi^0 \bar{K^0} $ is $1.62$, which is in good agreement with the experimental value of $1.64\pm0.29$. Similarly, the calculated branching ratio for $\Omega_c^0 \to \Xi^- \pi^+ $  is $0.15$, which aligns well with the experimental result of $0.25\pm0.060$.   For the decay $ \Omega_c^0 \to \Omega^- \pi^+ $, our predicted branching ratio is $1$, whereas the experimental value is $1.8$. In the case of $\Omega_c^0 \to \Omega^- K^+$, Our calculation yields $0.06$, which is consistent with the experimental upper limit of $<0.29$. For $ \Omega_c^0 \to \Omega^- \rho^+ $  we obtain a branching ratio of $1.72$, which is compatible with the experimental lower bound of $>1.3$. Overall, the theoretical prediction exhibits good agreement with the experimental measurement, reinforcing the validity of our approach. 

We analyze the color-allowed two-body nonleptonic decays of the bottom baryon $\Omega_b^-$ involving pseudoscalar and vector meson emission within the naïve factorization framework. The hadronic transition matrix element is factorized into a product of two independent matrix elements. The decay widths for both final states are expressed in terms of the parent and daughter baryon masses. For numerical evaluation, form factors are adopted from the light-front quark model, while CKM matrix elements, meson decay constants, and the $\Omega_b^-$ lifetime are taken from current experimental data. The decay mode $\Omega_b^- \to \Omega_c^0\pi^-$ has been observed experimentally; however, precise measurements of its decay width and corresponding branching ratio remain unavailable. Consequently, it becomes essential to compare our predicted branching ratios with those obtained from other theoretical frameworks to assess the degree of model dependence in these decay processes. 
In Table \ref{tab:Comparison}, we present a comparative analysis of our predicted branching ratios with various quark model predictions. The predicted branching ratios span a wide range—from $0.02$ to $12.2$—indicating notable variation across decay channels. This is consistent with the range of $0.02$ to $13.5$ reported in Ref.~\cite{Li:2021kfb}, supporting the overall agreement with existing theoretical models.
For the decay $\Omega_b^- \to \Omega_c^0\pi^-$, our predicted value is approximately half of that reported in Refs.\cite{Cheng:1996cs, Ivanov:1997hi, Ivanov:1997ra, Zhao:2018zcb}, but nearly twice that predicted in Refs.\cite{Gutsche:2018utw, Chua:2019yqh}, highlighting the model dependence for this channel.
In contrast, our results for $\Omega_b^-\to\Omega_c^0K^-$, $\Omega_b^-\to\Omega_c^0D^-$, $\Omega_b^-\to\Omega_c^0K^{*-}$, and $\Omega_b^-\to\Omega_c^0D^{*-}$ show close agreement with Refs.~\cite{Li:2021kfb, Zhao:2018zcb}, suggesting these modes are relatively insensitive to the choice of theoretical framework.
The decay $\Omega_b^- \to \Omega_c^0 D_s^-$ consistently emerges as the most favorable channel, with the highest branching ratio in our analysis, in agreement with other theoretical predictions.
Our calculated branching ratios are in good agreement with those reported in Ref.~\cite{Li:2021kfb}. However, substantial deviations from other theoretical approaches are evident, underscoring the importance of future experimental validation. In particular, confirmation of any one of the predicted decay modes would provide critical insight into the reliability of the various models used in describing non-leptonic bottom baryon decays.

Our approach successfully predicts experimental data by fixing model parameters (\(\lambda\), \(V_0\), and \(\sigma\)), demonstrating its potential for broader application. In our previous study \cite{Patel:2024wfo}, we applied this method to light baryons with extensive data. Here, we extend it to singly heavy baryons using the Independent Quark Model in a relativistic framework with a Martin-like potential. Our results not only align with experimental observations but also provide reliable predictions for future discoveries. Given this success, the approach can be applied to other well-studied baryons, allowing parameter correlations with mass and enabling predictions for baryons with limited or no experimental data.
\\
\begin{acknowledgments}
Ms. Rameshri Patel acknowledges the financial support from the University Grants Commission (UGC-India) under the Savitribai Jyotirao Phule Single Girl Child Fellowship (SJSGC) scheme, Ref. No.(F. No. 82-7/2022(SA-III)). We would like to express our gratitude to Prof. P.C. Vinodkumar for the insightful knowledge he imparted throughout this work.
\end{acknowledgments}


\end{document}